# LCA and energy efficiency in buildings: mapping more than twenty years of research

Asdrubali, F., Fronzetti Colladon, A., Segneri, L., & Gandola, D.M.

**LCA and energy efficiency in buildings: mapping more than twenty years of research**

**F. Asdrubali [a], A. Fronzetti Colladon [b], L. Segneri [b], D.M. Gandola [a]**

a. Department of International Human and Social Sciences, Perugia Foreigners' University, Piazza B. Fortebraccio 4, 06123 Perugia, Italy

b. Department of Engineering, University of Perugia, Via G. Duranti 93, 06125 Perugia, Italy

**Abstract**

Research on Life Cycle Assessment (LCA) is being conducted in various sectors, from analyzing building materials and components to comprehensive evaluations of entire structures. However, reviews of the existing literature have been unable to provide a comprehensive overview of research in this field, leaving scholars without a definitive guideline for future investigations. This paper aims to fill this gap, mapping more than twenty years of research. Using an innovative methodology that combines social network analysis and text mining, the paper examined 8024 scientific abstracts. The authors identified seven key thematic groups, building and sustainability clusters (BSCs). To assess their significance in the broader discourse on building and sustainability, the semantic brand score (SBS) indicator was applied. Additionally, building and sustainability trends were tracked, focusing on the LCA concept. The major research topics mainly relate to building materials and energy efficiency. In addition to presenting an innovative approach to reviewing extensive literature domains, the article also provides insights into emerging and underdeveloped themes, outlining crucial future research directions.

**Highlights:1**

- Social Network Analysis and Text Mining to analyze 8024 abstracts on LCA and Building
- The semantic brand score measures the semantic importance of seven issues related to Building and Sustainability
- Limited publications assessing the social impact of LCA in buildings
- Materials are the most central issue in the literature on sustainable buildings

## List of abbreviations

| | Nomenclature | Unit |
|---|---|---|
| **SBS** | Semantic brand score | - |
| **$CO_2$** | Carbon dioxide | |
| **LCA** | Life Cycle Assessment | - |
| **BSCs** | Building and sustainability clusters | - |
| **BMC** | Building materials and component | - |
| **ISO** | International Organization for Standardization | - |
| **RA** | Review articles | - |
| **MA** | Methodological aspects | - |
| **DP** | Design procedures | - |
| **SM** | Sustainable materials | - |
| **CS** | Case studies | - |
| **TF-IDF** | Term-Frequency Inverse-Document-Frequency | - |
| **PR** | Prevalence | - |
| **DI** | Diversity | - |
| **CO** | Connectivity | - |
| **$\overline{PR}$** | Mean value of Prevalence | - |
| **$\overline{DI}$** | Mean value of Diversity | - |
| **$\overline{CO}$** | Mean value of Connectivity | - |
| **STD** | Standard deviation | |
| **BI** | Business Intelligence | - |
| **ELCA** | Environmental Life Cycle Assessment | - |
| **LCC** | Life Cycle Costing | - |
| **S-LCA** | Social Life Cycle Assessment | - |
| **UN** | United Nations | - |

## 1. Introduction

According to the 2022 United Nations "Global Status Report for Buildings and Construction" [1], the decarbonization of building stock is too slow and "not on track" to reach the goals of the Paris

Agreement. The world's energy needs and greenhouse gas emissions from the building sector have increased by approximately 10% in the years 2010-2021, and there are concerns about resource consumption due to construction materials in the next decades, especially in fast-growing developing countries.

Decarbonizing the building sector is urgent and may take place only under a common framework of procedures and evaluation tools to quantify and reduce energy consumption and $CO_2$ emissions. To this extent, the Life Cycle Assessment approach considers all the various stages of construction processes and is the most comprehensive methodology. LCA is widely known as a methodology for investigating the energy and environmental impacts of products and processes from "cradle to grave". Building materials and components (BMC) play an important role in the environmental performance of a building and affect various stages of the life cycle. Therefore, LCA can bring environmental consciousness into the selection of sustainable building materials and components [2].

The LCA approach has been standardized by ISO [3,4] and can be used to provide a basis for assessing potential improvements for the eco-profile of products and, therefore, to design more efficient and environmentally friendly complex products, as entire buildings and building materials and components. Various LCA-based benchmarks have been developed as part of regulations, labeling systems, sustainability rating tools, and research studies, as well as many tools have been developed to assess the sustainability of buildings. Furthermore, many environmental labels have been proposed to evaluate the properties and impacts of construction materials and components [5]. LCA studies applied to buildings have gained importance in the scientific literature recently. However, many questions still need to be solved since buildings are very complex systems, which implies a very accurate and critical approach.

Buildings are very complex "products" with a long lifetime, made of many different materials, components, and technological systems, often assembled with limited industrialization so that each building is different from the others: this implies methodological peculiarities in the definition of the functional unit, the system boundaries, the cut-off criteria, the allocation procedures. An important parameter is the "embodied" energy or "grey" energy, which is the measure of the energy used in the BMC production, on-site construction, and end-of-life; this parameter has recently gained importance, especially for Nearly Zero Energy Buildings, for which there is a real risk of shifting the impacts from the operational phase to the construction and end of life phases [6]. The LCA approach can be very useful for new and existing buildings in identifying the 'hot spots', that is, the most impacting phases, processes, components and materials. Identification allows for making more sustainable and conscious choices.

The literature shows many studies highlighting the role of LCA as a decision-making support tool in selecting low-impact BMC and designing green buildings. An emerging trend in the literature on LCA is dedicated to low-energy buildings, in which the goal of reducing the operational energy implies an increase in embodied energy due to additional materials and components used in the building shell and technical equipment for energy efficiency measures [7].

Starting from the initial focus on the environmental aspects, LCA has recently incorporated economic and social aspects [8,9]; in the future, LCA will be crucial for supporting all these aspects in the building sector, not only for the definition of strategies for impact reduction but also to measure the effectiveness of policy interventions, scaling up the benefits and impacts from the building level to the city and regional level.

Recent research has also brought to light the significant environmental benefits that can be achieved through the implementation of circular economy practices [10–14]. In comparison to traditional recycling methods, the design and construction of buildings with reusability in mind have been shown to reduce greenhouse gas emissions by as much as 88 percent, while also improving various environmental indicators. One promising approach involves creating modular buildings, specifically designed for disassembly and the reuse of components, presenting a tangible opportunity to mitigate greenhouse gas emissions and other environmental impacts [10]. Furthermore, a recent study explored the use of industrial by-products such as fly ash, bottom ash, and recycled concrete as alternatives to natural aggregates, taking into account factors such as cost, environmental impact, and energy consumption [13]. The findings revealed no superior material across all categories, as each material possesses unique advantages and disadvantages [11]. As such, there is a pressing need for comprehensive studies in the field that offer a standardized set of emission factors derived from life cycle assessments conducted in accordance with ISO 14040 standards, encompassing a wide range of recycled materials [14]. This standardized approach facilitates a clear comparison of the environmental performance of different materials, thereby enhancing the accuracy of assessments related to reuse options. Moreover, using recycled concrete has emerged as a viable solution to address the scarcity of natural aggregates in certain regions, such as Switzerland [12]. Research has delved into the environmental and economic efficacy of incorporating recycled materials into road pavements, demonstrating a reduction in key environmental impacts with an increase in the proportion of recycled material. These findings underscore the potential of recycled materials to mitigate environmental harm and offer cost-effective solutions for sustainable infrastructure development.

The topic of LCA applied to buildings and building materials is relatively recent. Table 1 provides an overview of the top 20 most cited papers on applying LCA in the building sector. Of note is the origin of authors from all over the world, with a prevalence of Europe. However, significant contributions also come from emerging countries such as China, India, and Brazil. This shows that the LCA of buildings is an important research topic for many scholars. Furthermore, the heterogeneity of applications demonstrates the versatility of the topic, which is studied from different points of view, such as general review articles (RA), methodological aspects (MA), design procedures (DP), sustainable materials (SM), and case studies (CS).

**Table 1.** High-Impact Papers on "LCA" and "Construction" Published in the Last 15 Years

| Paper Title | Journal | Point of view | Year | Citations | Percentile | Country |
|---|---|---|---|---|---|---|
| Life cycle energy analysis of buildings: An overview [15] | Energy and Buildings | RA | 2009 | 919 | 98 | India |
| Sustainability in the construction industry: A review of recent developments based on LCA [16] | Construction and Building Materials | MA | 2008 | 904 | 99 | Spain |
| Eco–efficient cements: Potential economically viable solutions for a low-CO2 cement-based materials industry [17] | Cement and Concrete Research | SM | 2018 | 874 | 99 | Switzerland |
| Life cycle assessment (LCA) and life cycle energy analysis (LCEA) of buildings and the building sector: A review [18] | Renewable and Sustainable Energy Reviews | RA | 2014 | 853 | 98 | Spain |
| Life cycle assessment of building materials: Comparative analysis of energy and envir onmental impacts and evaluation of the eco-efficiency improvement potential [19] | Building and Environment | MA | 2010 | 813 | 99 | Spain |
| Sustainable construction—The role of environmental assessment tools [20] | Journal of Environmental Management | MA | 2007 | 789 | 99 | Australia |
| Life-Cycle Assessment and the Environmental Impact of Buildings: A Review [21] | Sustainability | RA | 2009 | 482 | 97 | United Kingdom |
| Embodied GHG emissions of buildings – The hidden challenge for effective climate change mitigation [22] | Applied Energy | RA | 2019 | 319 | 99 | Austria |
| Circular economy in the construction industry: A systematic literature review [23] | Journal of Cleaner Production | RA | 2020 | 187 | 97 | Brazil |
| Integration of LCA and LCC analysis within a BIM-based environment [24] | Automation in Construction | DP | 2019 | 136 | 99 | Belgium |
| A review of the life cycle assessment of buildings using a systematic approach [25] | Building and Environment | RA | 2019 | 113 | 95 | United States |
| Modelling of energy consumption and carbon emission from the building construction sector in China, a process-based LCA approach [26] | Energy Policy | MA | 2019 | 100 | 97 | China |
| Integrating building information modelling and life cycle assessment in the early and detailed building design stages [27] | Building and Environment | DP | 2019 | 92 | 98 | Canada |
| Construction solutions for energy efficient single-family house based on its life cycle multi-criteria analysis: A case study [28] | Journal of Cleaner Production | CS | 2016 | 95 | 97 | Lithuania |
| Benchmarks for environmental impact of housing in Europe: Definition of archetypes | Building and Environment | MA | 2018 | 90 | 97 | Italy |
| and LCA of the residential building stock [29] | | | | | | |
| Critical consideration of buildings' environmental impact assessment towards adoption of circular economy: An analytical review [30] | Journal of Cleaner Production | RA | 2018 | 89 | 95 | China |
| Embodied carbon emissions of office building: A case study of China's 78 office buildings [31] | Building and Environment | CS | 2016 | 98 | 97 | China |
| Dynamic LCA framework for environmental impact assessment of buildings [32] | Energy and Buildings | MA | 2017 | 87 | 95 | China |
| Exergy analysis and life cycle assessment of solar heating and cooling systems in the building environment [33] | Journal of Cleaner Production | MA | 2012 | 87 | 94 | Greece |
| Evaluation of whole life cycle assessment for heritage buildings in Australia [34] | Building and Environment | MA | 2012 | 82 | 94 | Australia |
| Life cycle assessment of recycled concretes: A case study in southern Italy [35] | Science of the Total Environment | CS | 2018 | 78 | 93 | Italy |
| A comparative Life Cycle Assessment of external wall compositions for cleaner construction solutions in buildings [36] | Journal of Cleaner Production | MA | 2016 | 78 | 93 | Italy |

The diverse articles showcased here underscores the global impact of research on Life Cycle Assessment (LCA) and building sustainability, with substantial contributions originating from multiple continents. An examination of citations and percentages further validates the profound influence these studies have had on the field of sustainability.

It can be argued that LCA applied to buildings has been an emerging and global topic in the scientific literature for the last 15 -20 years, and it deserves a detailed analysis to understand its relevance and future trends better. Accordingly, this work aims to present an in-depth analysis of the importance of the concept of LCA applied to buildings in the scientific literature over the past two decades.

Due to the extensive body of research on the topic, the authors used an innovative literature review approach based on the integration of social network analysis and text mining. By applying this methodology to the titles and abstracts of the selected articles, they identified seven key thematic groups, called building and sustainability clusters (BSCs). To gauge the importance of each cluster within the broader discourse on building and sustainability, they have used the semantic brand score (SBS) indicator [37]. Additionally, this article explores the interconnections between different clusters and monitored trends over time, focusing on the LCA concept. In this way, it allows to understand the main issues conveyed in each document and how they relate to other topics in the literature.

Using this methodology, this paper gains valuable insight into the academic discussion on LCA and the building sector and uncovers some intricate connections between LCA and other relevant topics. This analysis contributes to the literature on LCA in buildings and provides a deeper understanding of the interplay between its main related topics, thus discussing past, present, and future research.

The contribution of this research is inherent in the methodology used to review this literary strand. Indeed, the semi-supervised methodology enables the analysis of a vast volume of abstracts spanning a broad time frame in the sustainability and building literature, encompassing 1996 to 2022. This differentiates this study from other reviews focused on narrower time frames. For instance, some studies analyzed 15 years but omitted data before 2000 [38], while others covered fewer years and involved a smaller number of papers [39,40]. To the best of the authors' knowledge, only Kaklauskas et al. (2021a) included in their analysis scientific publications that occurred during the years affected by Covid-19.

This literature review also addresses the objectives outlined in Agenda 2030, also known as the Sustainable Development Goals (SDGs), specifically focusing on Goals 7, 11, 12, and 13. The utilization of life cycle assessments (LCA) in building materials is crucial for promoting the adoption of more efficient and sustainable energy solutions, in alignment with Goal 7. LCA research plays a key role in enhancing the sustainability of cities by evaluating the environmental impact of urban infrastructure and facilitating improvements, as outlined in Goal 11. Furthermore, integrating LCA practices in building design contributes to reducing environmental impacts associated with building materials, thereby promoting responsible production and consumption practices in line with Goal 12. Additionally, incorporating LCA methodologies in building design aids in combating climate change by decreasing $CO_2$ emissions and enhancing the energy efficiency of buildings, as highlighted in Goal 13.

The market is witnessing a growing emphasis on sustainability and materials lifecycle analysis, prompting industry professionals to prioritize environmental considerations alongside technical and economic factors. Tools like SimaPro have been developed to objectively quantify environmental benefits between different materials, facilitating informed decision-making. Furthermore, the introduction of international standards, such as the ISOs mentioned above, aims to safeguard the market and uphold the sustainability of production processes and construction materials. These advancements underscore the industry's commitment to sustainable practices and the importance of integrating environmental considerations into decision-making processes.

The structure of the paper is as follows. Section 2 presents the methodology used. Section 3 describes the main results of the analysis from both the temporal and spatial points of view. Section 4 draws the conclusions.

## 2. Methodology

This work presents an innovative literature review approach to obtain a semi-automated and comprehensive understanding of building life cycle assessment research. The authors first selected abstracts as described in Section 2.1, and then, using a semantic network approach, analyzed them to explore their primary topics, trends, and associations. Figure 1 illustrates the methodological flow followed in this literature review.

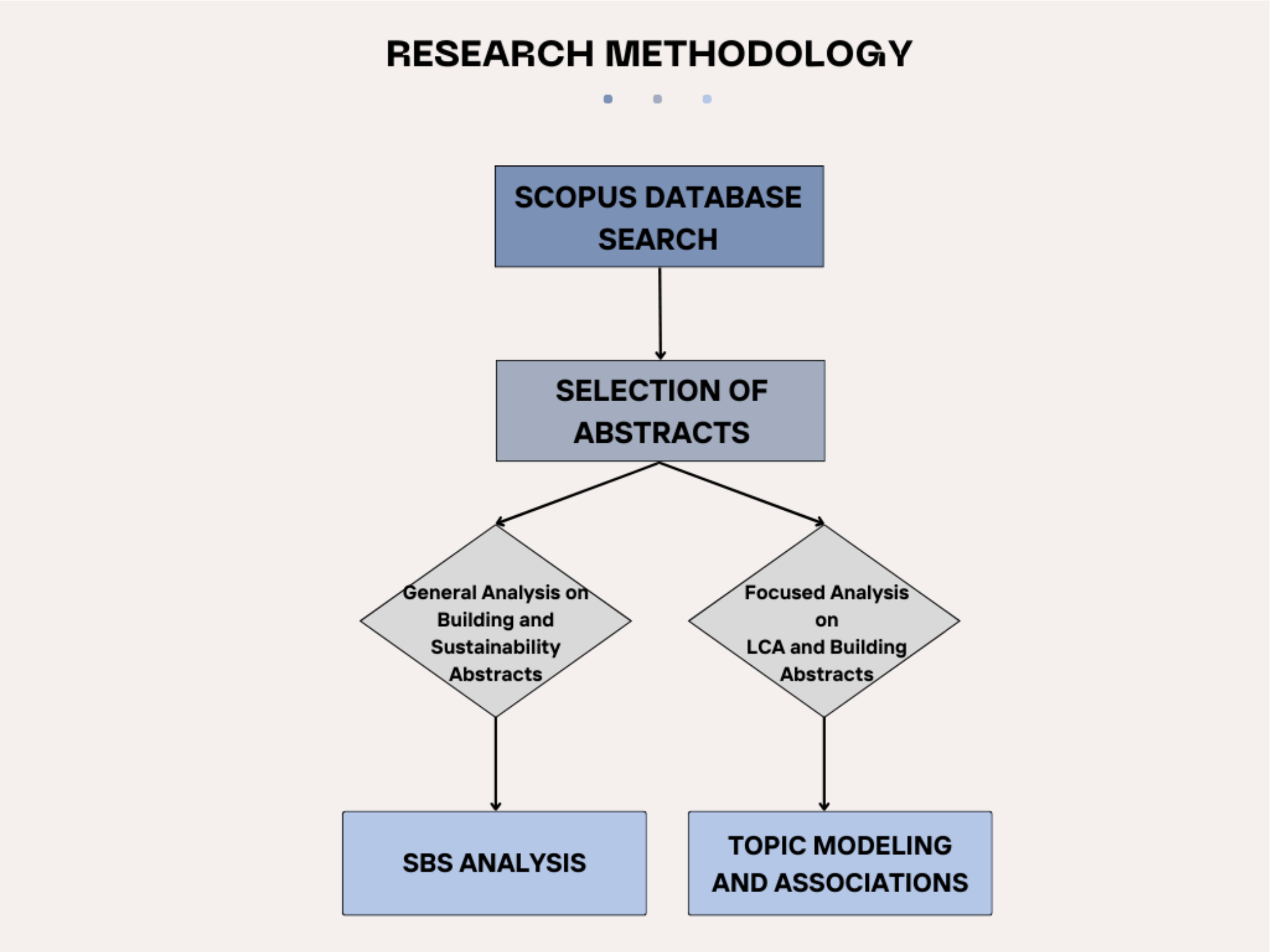

**Figure 1.** Graphical representation of the implemented methodology.

### *2.1. Data collection*

To ensure the relevance and homogeneity of the selected papers, the authors implemented a filtering process on journal articles published between 1996 and December 2022. Scopus has been the reference database because of its comprehensive coverage of academic literature, including abstracts and citations from many academic sources [42].

Since the data collection was conducted in mid-2023, the authors excluded all works published after December 2022. Additionally, they retained only papers written in English and published in scientific journals. Lastly, taking as reference two pivotal works in the literature review on sustainability and building [43,44], they constructed a query that included all articles with keywords that contained the words building and sustainability. Using this criterion a first dataset, “Building and Sustainability”, consisting of 4744 papers, was created.

In addition, the authors created a second database of papers, “LCA and Building” dataset, including work focused on applying life cycle assessment in the context of buildings. It comes from filtering the same original dataset, but this time, considering all articles that included the words life cycle assessment or its acronym “LCA” and the word building. As a result, a dataset composed of 3270 papers was obtained.

### *2.2 Text Mining and Semantic Network Analysis*

This research applied an innovative approach combining text mining and social network analysis methods to analyze the abstracts of previously selected papers and provide a comprehensive overview of their main topics. As detailed in Section 2.2.1, the authors first identified a set of clusters representing the main themes discussed in the literature. Then, using the semantic brand score (SBS) indicator [37] the authors discuss its semantic importance over time, also analyzing their image and associations (Section 2.2.2). For the analyses presented in this study, the SBS BI web application [45] was used.

#### *2.2.1 Building and sustainability clusters selection*

After the data collection phase, the authors conducted a semi-supervised selection of the main topics and identified the words that better describe them. First, the main keywords were extracted, employing a Term-Frequency Inverse-Document-Frequency (TF-IDF) approach [46]. Subsequently, two experts were convened to analyze this initial set of keywords. The goal was to select and categorize the keywords into meaningful groups, each representing a specific concept or theme. The experts were also supported by the Lexicon Augmenter tool in the SBS BI app, which helps to expand an initial word list by identifying synonyms, hyponyms, hypernyms, and related terms. For example, if the input term is “home”, the tool suggests “cottage” as a hyponym and “residence” or “domicile” as hypernyms. The tool exploits the Wordnet lexical database [47] and pre-trained word embedding models [48,49] to improve keyword selection. Combining the experience and domain knowledge of the experts, insights from past research, and the information provided by SBS BI, the authors obtained seven groups of keywords, the building and sustainability clusters (BSCs). The complete list of these clusters and examples of their most important words can be found in Table 2.

**Table 2.** Overview of Building and Sustainability Clusters, Keywords, and Descriptions

| BSCs | Keywords Examples | Description |
|---|---|---|
| Renewables | renewable energy, solar energy, geothermal energy, hydroelectric energy, wind energy, biomass energy | Terms related to sustainable energy sources |
| Efficiency | energy efficiency, refurbish, retrofit | Terms used to identify the concept of efficiency in the construction process |
| Materials | brick, insulation, portland cement, alkali, limestone, asphalt, sand, stone, steel, glass, wood | Terms related to various materials commonly used in construction and building processes |
| Components | composite, roof, green roof, cool roofs, façade, window | Terms related to building elements and features that aim to improve sustainability and energy efficiency |
| Energy Technologies | heat pump, solar panels, solar panels, photovoltaic, micro-wind, boiler | Terms associated with the technology responsible for generating, converting, storing, and distributing various forms of energy |
| LCA | life cycle assessment, life cycle energy, lce, life cycle cost | Terms related to the building life cycle assessment approach, which is the procedure for assessing the energy and environmental loads related to a building process |
| Saving | economic benefits, saving, payback time, investment | Terms related to economic aspects and advantages associated with various initiatives and investments |

It categorizes a collection of terms related to the construction and building industry, specifically focusing on sustainable practices, energy efficiency, materials, components, and economic considerations. The cluster *renewables* includes terms related to sustainable energy sources, such as renewable energy, solar energy, geothermal energy, hydroelectric energy, wind energy, and biomass energy. *Efficiency* encompasses terms used to describe and measure efficiency in construction processes, including concepts such as energy efficiency, refurbishment, and retrofitting. The *materials* cluster involves terms associated with various construction materials commonly used in building processes. *Components* includes terms related to specific building elements and features designed to improve sustainability and energy efficiency. *Energy technologies* refer to terms related to technology used for generating, converting, storing, and distributing various forms of energy. The *LCA* cluster focuses on terms associated with the life cycle assessment approach in building construction, while *saving* addresses economic aspects and benefits associated with various initiatives and investments in the construction industry.

### *2.2.2. Measurement of BSCs' importance with the semantic brand score*

The semantic brand score (SBS) is a metric that enables the measurement of the significance of words and concepts within large text corpora. Before it can be calculated, the textual data must be preprocessed and transformed into a network of words. Text preprocessing includes removing punctuation, stop words and special characters, converting text to lowercase, computing bigrams, and extracting stems [50]. These steps simplify the language, retain meaningful words, and serve the creation of a network where nodes represent words, and arcs represent their cooccurrences. In particular, arcs are weighted based on the co-occurrence frequency of two connected terms, with higher weights indicating a stronger connection. Once the pre-processing phase is complete, a network of words is created for each year, with the keywords representing each BSC clustered into single nodes. Subsequently, the authors calculated the semantic importance of each cluster using the SBS indicator – which is made of three dimensions, namely prevalence (PR), diversity (DI), and connectivity (CO). The sum of these three standardized dimensions measures the importance of a BSC, as given by the formula 1:

$$SBS(i) = \frac{PR(i) - \overline{PR}}{std(PR)} + \frac{DI(i) - \overline{DI}}{std(DI)} + \frac{CO(i) - \overline{CO}}{std(CO)} \quad (1)$$

where $i$ denotes a BSC, $PR(i)$ is its prevalence, $DI(i)$ its diversity, and $CO(i)$ its connectivity. Delving into the specifics of the metrics, the first dimension is the prevalence. It represents the frequency of a BSC within the corpus, that is, how often that specific research cluster is discussed in the abstracts. The more scientific articles that mention a particular topic, the more the research community becomes aware of it, potentially leading to an increase in future studies. This construct shares partial relevance with brand awareness, which posits that repeating a specific word in the text enhances its memorability and recognition [51].

Diversity measures the heterogeneity and uniqueness of words associated with a BSC, partly related to the concept of brand image [51]. It is calculated through the *distinctiveness centrality metric* [52]. The calculation of this metric requires the transformation of the texts into networks, as described by the formula 2.

$$DI(i) = \sum_{\substack{j=1 \\ j\neq i}}^{n} \log_{10} \frac{(n-1)}{g_j} I(w_{ij} > 0) \tag{2}$$

Equation 2 measures the distinctiveness of a BSC, i.e., node $i$ in the network. In the formula 2, $n$ is the total number of nodes, $g_j$ is the degree of node $j$, and $I(w_{ij} > 0)$ is an indicator function that is equal to 1 when the edge connecting nodes $i$ and $j$ exists – i.e., when $w_{ij} > 0$ – and is equal to 0 when this edge is missing. Diversity grows with the number of textual associations, attributing more importance to those that are distinctive and less common.

The last dimension is connectivity. It is operationalized through *weighted betweenness centrality* [53], which indicates a node's (word or cluster) ability to link topics or other words that are not directly connected. Connectivity is calculated using the formula 3:

$$C(i) = \sum_{j<k} \frac{s_{ik(i)}}{s_{jk}} \tag{3}$$

where $s_{jk}$ equals the number of the shortest network paths linking the generic pair of nodes $j$ and $k$, and $s_{jk}(i)$ is equal to the number of those paths containing node $i$. This measure can be intended as a proxy of the brokerage power of a word or cluster, that is, its ability to connect different parts of the overall discourse.

## 3. Results

A total of 4744 abstracts were extracted from the initial search query related to research on sustainable buildings. In Figure 2, it can be observed the number of articles published since 1996. An upward trend can be seen, especially from 2007, with occasional fluctuations. Since 2001, there has been a gradual and consistent increase in the yearly number of publications. In particular, the number of publications increased significantly from 2007 to 2013. It remained relatively stable in the following years until 2015, with slight variations. The highest number of publications was observed in 2021, with 643 published papers. However, between 2016 and 2022, a steady annual increase indicates a sustained growth in research output in this field.

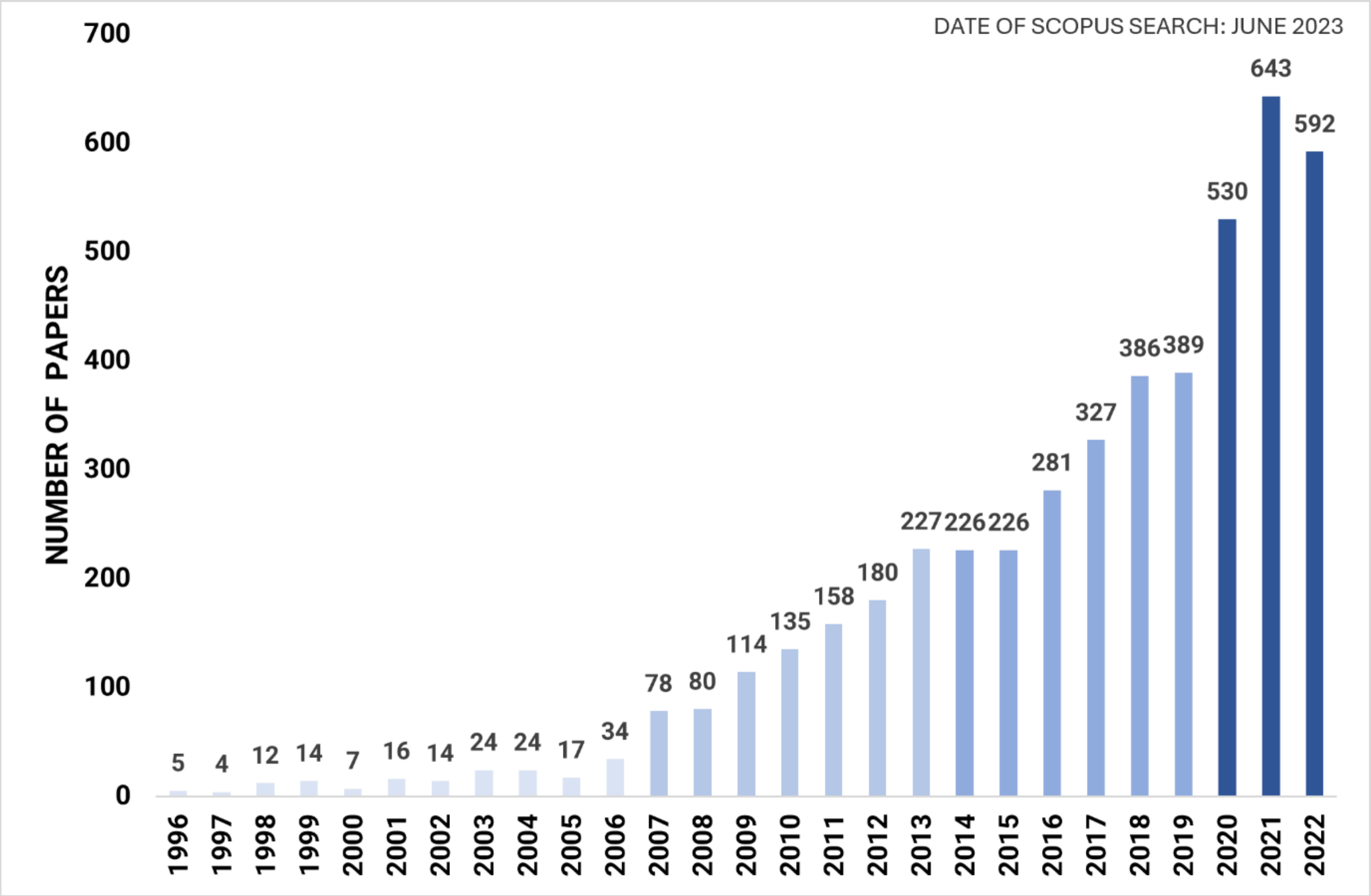

**Figure 2**. Number of papers found searching for "building" and "sustainability" in Scopus

### *3.1 Evolution of Building and Sustainability Literature*

Looking at Figure 2, four evolutionary phases of the building and sustainability literature were identified , highlighted with different tones of blue. An initial development phase from 1996 to 2006, a rapid development phase from 2007 to 2013, a differentiation phase of relevant issues between 2014 and 2019, and a peak phase in the number of publications observed from 2020 to 2022.

#### *3.1.1. 1996-2006: Phase of initial development*

The number of publications remained relatively low and stable throughout the reporting period, with a slight increase in 2006. On average, there were about 12 publications per year until 2005. However, in 2006, this number increased significantly to 34 publications. This indicates an early stage of scientific production in the general field of building and sustainability. The upward trend is observed primarily in the latter part of the period, as confirmed by several studies [44,54,55]. Looking at the SBS trends in Figure 3, it can be noted that *efficiency* and *savings* are of the highest semantic importance. The increasing interest in *efficiency* leads us to hypothesize that scientific attention is beginning to focus on improving energy efficiency in building performance. This initial attention is also highlighted in a comprehensive review on green building [56], where the authors emphasize the growing importance of words such as *"energy efficiency"* and *"performance"*. Furthermore, Figure 3 shows a growing trend in the *savings* SBS.

This result suggests that researchers are exploring strategies and methods to save resources in terms of time, materials, and costs to improve construction processes in the building sector. This reasoning is supported by the findings of He et al. [57], who analyzed sustainable building renovation. The authors highlight that since the early 2000s, the scientific community has focused on studying various tools to streamline time and resource efficiency in building renovation.

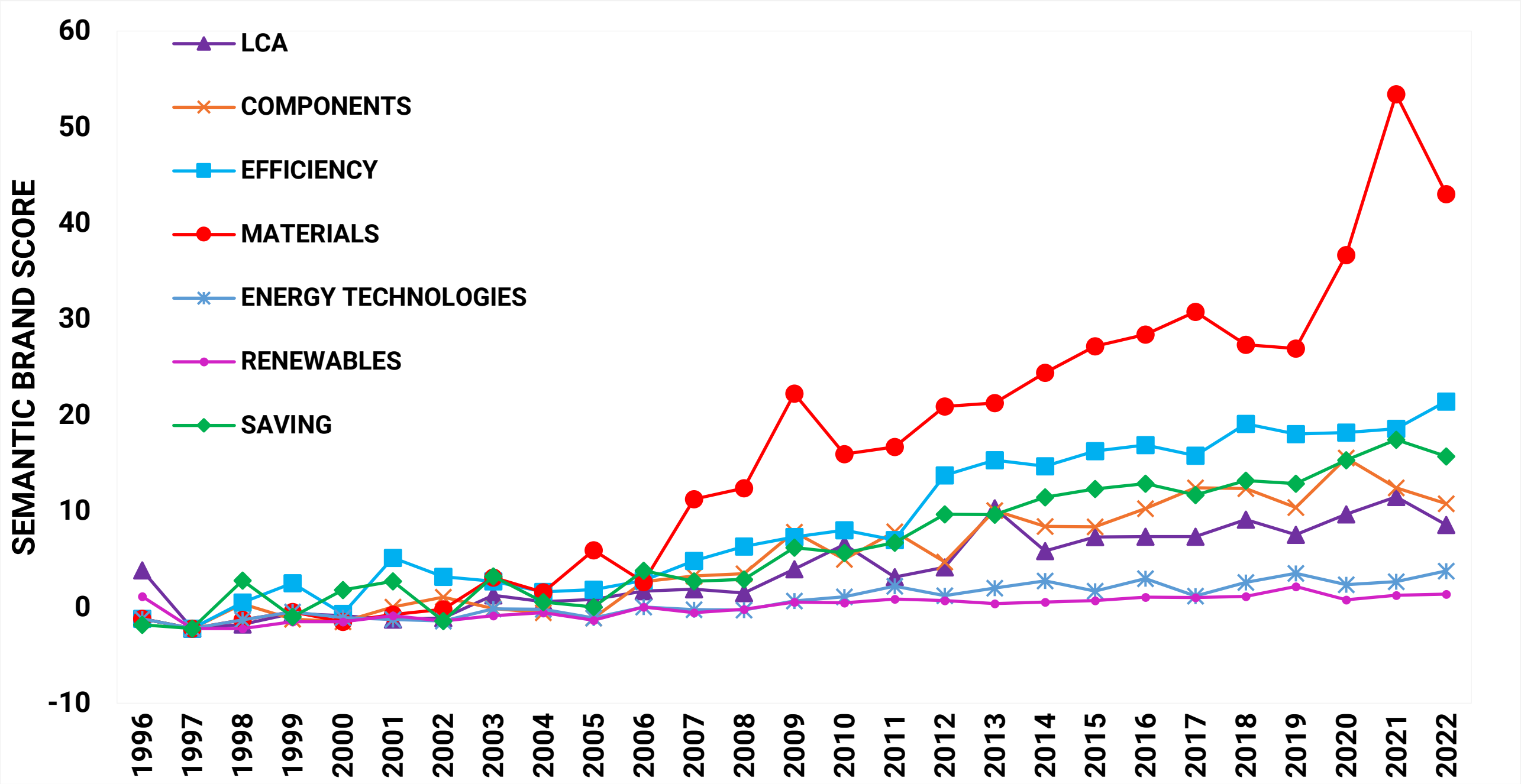

**Figure 3.** SBS time series of the BSCs

### *3.1.2. 2007-2013: The phase of rapid development*

As shown in previous studies, the number of publications has increased steadily over this period [44,58,59]. A significant increase in the semantic importance of *materials*, *efficiency*, *lca,* and *components* can also be observed. The initial attention towards the LCA cluster can be attributed to the publication in 2006 of ISO 14040 [3]. It establishes the principles and framework for life cycle assessment. Furthermore, during the entire period, numerous authors have highlighted the initial interest shown by the scientific community in various aspects of life cycle assessment, such as the quantitative analysis of embodied energy [60] or the role of carbon [61]. Subsequently, the LCA trend demonstrates significant growth, reaching its peak in 2013, as Crippa et al. (2020) also show. Moreover, the increasing SBS trend of *efficiency* can be compared with the parallel growth observed in *materials* and *components*. This indicates the desire of researchers to explore approaches aimed at improving the energy efficiency of buildings by optimizing building materials and components [61].

### *3.1.3. 2014-2019: The differentiation phase*

Figure 2 shows a consistent upward trend in the number of publications between 2014 and 2019, accounting for more than 60% of total publications from 1996 to 2019. The prevalence of publications during this period of time was also observed by Lima et al. [62] in their review of the sustainability and civil construction literature and by Kiani Mavi et al. [63] in their work on sustainability in construction projects. Of interest is the surge in publications since 2016. This could be attributed to the growing interest in sustainability within the construction sector, spurred by increased institutional attention [8]. This increase in publications could be attributed to the Paris Agreement, adopted by 196 Parties at the U.N. Climate Change Conference 2015. The agreement emphasizes limiting the rise in global temperature and transitioning to clean energy. Additionally, authors such as Kiani Mavi et al. [64] and Backes and Traverso [8] argued that the 2030 Agenda for Sustainable Development, adopted by all United Nations Member States in 2015, could impact the growing interest in publications in the field of sustainability and construction. The peak of SBS importance of the *efficiency* cluster in 2018 could reflect scientific development in material and building component performance and energy optimization [56,65,66]. In these works, researchers explored the notion of efficiency in the energy sector and demonstrated how green buildings outperform those constructed using conventional techniques.

### *3.1.4. 2020-2022: The peak phase*

The number of publications in this period is significantly higher than before, with 1765 total articles. To explain this upward trend, two lines of reasoning were followed. According to Backes and Traverso [8], this increase can be attributed to the continued impact on the scientific community of initiatives such as the Paris Agreement and the 2030 Agenda for Sustainable Development, both introduced in 2015. However, this increase in publications could also be related to the Covid-19 pandemic. Kaklauskas et al.[67] highlighted how pandemic-induced demand, such as increased time spent indoors, has forced citizens and the scientific community to prioritize sustainability concerns in the building sector. Indeed, the restrictions resulting from the Covid-19 pandemic have profoundly transformed how individuals and communities live, interact, and work [67]. This translates into greater awareness of the importance of investing in environmental protection and resilience of the built environment, residences, offices, entertainment venues, public buildings, and various indoor settings [68,69]. For example, Wang et al. [70] explained how green buildings can decrease the likelihood of infections and mitigate the risk of cross-contamination. In addition, they can improve general well-being and contribute to the continuity of productive work environments during epidemics. Upon closer examination of the abstracts retrieved by the query, a clear pattern emerges in publications pertaining to Construction and Sustainability. There is a notable surge in publications in 2021, followed by a decline in 2022. Specifically, 25 abstracts from 2021 heavily emphasize the concept of COVID-19. In contrast, in the 2022 articles, the mention of this concept decreases, with only 16 abstracts referencing it. This trend indicates that in 2021, many papers where published in response to the impact of the pandemic. However, by 2022, the focus had diminished, likely due to the initial wave of research and discussion in 2021 adequately addressing the effects of the pandemic on this field.

### *3.2. A focus on LCA*

After exploring the global literature on sustainable building, this research extended the analysis with an in-depth exploration of the second set of 3270 abstracts obtained through the second search query, that is, those specifically focused on research about LCA in buildings. Specifically, the authors modeled the main topics of the discourse using the Louvain clustering algorithm [71] in the abstracts' semantic network. Subsequently, they selected the most relevant words for each topic based on their weighted degree and the proportion of internal and external links, following the methodology proposed by Fronzetti Colladon and Grippa [45]. In this way, six main discourse topics were identified, as presented in Figure 4. Each word cloud represents a specific topic, with the size of the words indicating their importance within that topic.

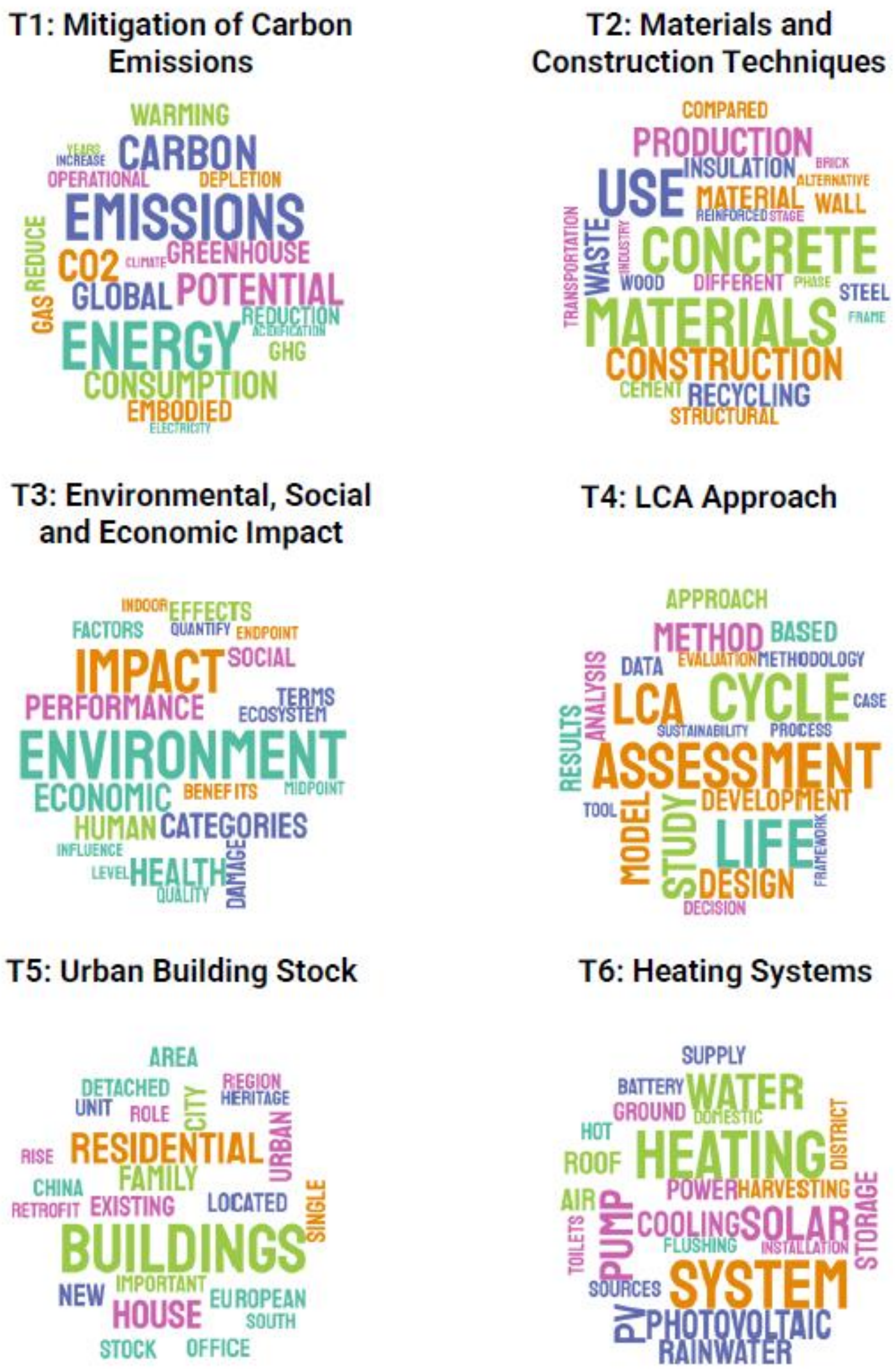

**Figure 4.** Detailed Breakdown of Main Discourse Topics in LCA and Building Research Abstracts

The first topic is related to emissions, the most important words being *"emission"*, *"carbon"*, *"consumption"*, *"co$_2$"*, and *"greenhouse"*. This suggests that academic work on LCA in buildings explores strategies to minimize carbon emissions and mitigate adverse environmental and climate effects. As supported by Anand and Amor [59], building LCA research has focused on improving energy efficiency and mitigating emissions during the operational phase [72]. This emphasis prompted extensive research and development endeavors aimed at enhancing the energy-efficient operation of buildings and, in turn, resulted in a transition of impacts towards the construction phase [73].

This shift in focus is supported by Topic 2, which pertains to "Materials and Construction Techniques". Due to advancements in research on the operational phase of buildings, there has been growing attention on the embodied energy of buildings [74–76]. Embodied energy refers to the total energy consumed throughout the entire life cycle of a building, from the extraction and *"production"* of *"materials"* to their *"transportation"*, *"construction"*, and disposal or *"recycling"*. It encompasses the energy required to manufacture and process construction materials, such as *"cement"* and *"steel"*, and the energy expended during the construction and assembly of the structural components. Interestingly, Topic 3, "Environmental, Social, and Economic Impact", brings attention to an additional facet of LCA and building research, specifically related to the broader concept of sustainable development. While there has been a growing recognition of the need to evaluate the environmental and economic performance of buildings, leading to the creation of various tools such as Environmental Life Cycle Assessment (ELCA) and Life Cycle Costing (LCC), the social dimension of the life cycle of buildings, known as Social Life Cycle Assessment (S-LCA), is a relatively recent field that is still in its early developmental stages [77–79]. This is also evident in the different dimensions of the words *"environmental", "economic",* and *"social"* within the word cloud of Topic 3.

A dedicated topic focuses on the evaluation of the life cycle assessment, and it covers technical aspects such as *"study design", "modeling", "methodology"*, data *"analysis"*, and *"decision-making"*.

In addition, a topic characterized by words such as *"urban", "building",* and *"stock"* was found. This suggests research attention to the application of LCA to building stocks, such as *"office", "residential buildings",* or *"single-family houses"*[80]. Additionally, the presence of the word *"retrofit"* is justified by the use of the urban building stock analysis literature to evaluate the impacts of retrofitting [81].

The last topic includes terms such as "*heating", "water", "solar energy", "pumps", "photovoltaic systems", "cooling", "rainwater harvesting",* and *"storage"*. This topic suggests a specific line of research on heating systems, with a particular focus on the water footprint of buildings, as confirmed in the work of Change et al. [82].

The heterogeneity of the semantic image of LCA is also presented in Figure 5, which provides an overview of the most distinctive and frequently associated words with LCA throughout the period (1996-2022).

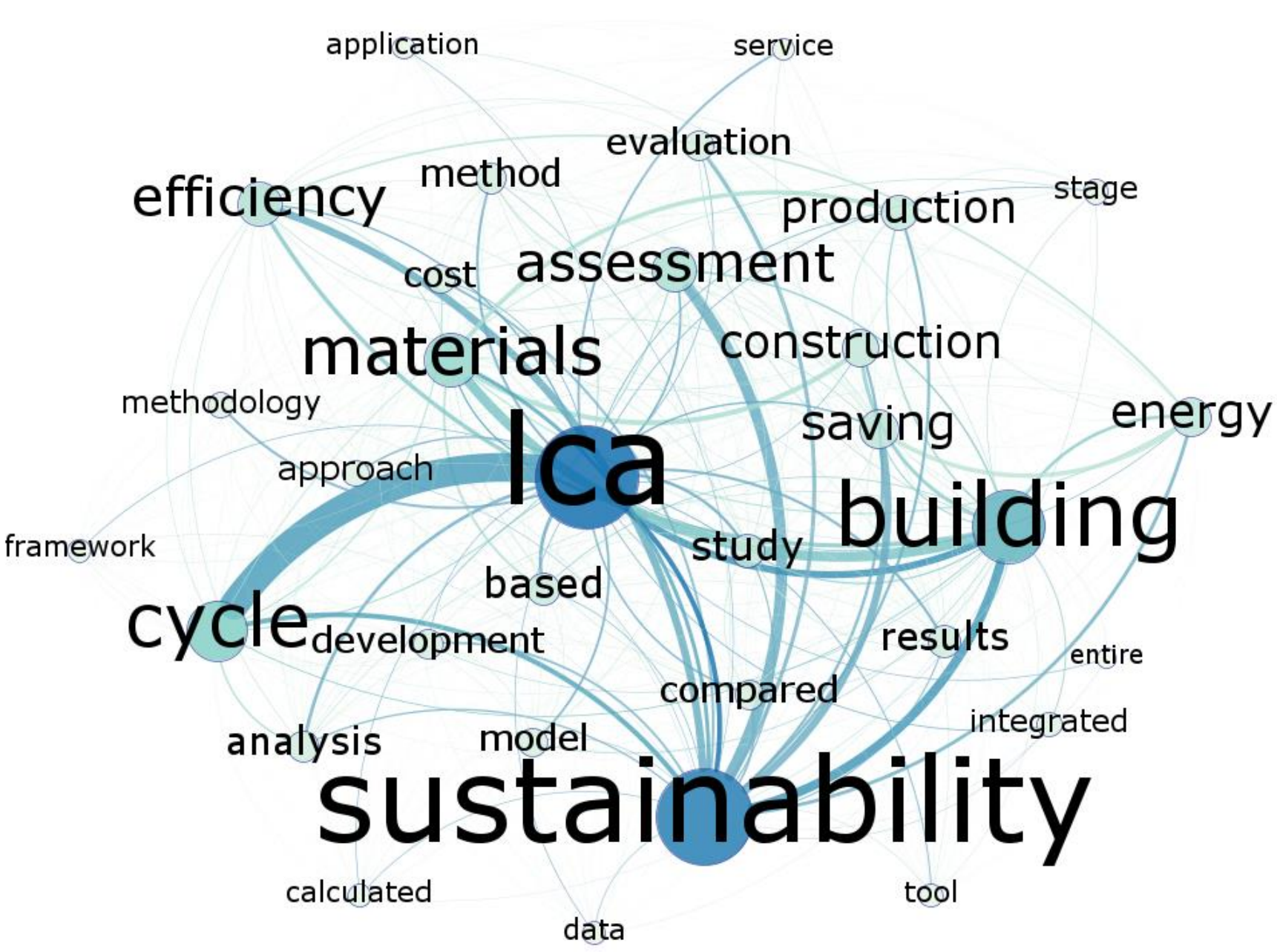

**Figure 5**. Top Associations with LCA in Research (1996-2022)

In line with what has been discussed on the main discourse topics, it can been observed that the semantic image of LCA is characterized by most of the words describing its operational nature, such as *"approach", "methodology", "method", "tool", "framework", "study"* and *"evaluation"*. These words suggest the presence of a research stream that uses LCA as an assessment tool for buildings and subsequent green certification. Numerous certifications are available for assessing buildings, generally considered more user-friendly than performing a complete LCA. However, it is essential to note that achieving a higher score through certification does not necessarily indicate a lower environmental impact during the entire life cycle of a building. This highlights the importance of integrating building certification tools with LCA [2,83]. For example, a study [84] emphasizes the need to integrate Life Cycle Assessment (LCA) with green building certifications to ensure a holistic approach to sustainability in the construction industry. Another article [85] focuses on the application of Building Information Modeling (BIM) integrated with LCA to compare the environmental impacts of wooden houses versus concrete masonry houses.  Furthermore, the remaining words describe various fields of application and indicate the direction scientific research has taken in using LCA in the construction sector. Words such as *"materials", "construction", "production", "energy", "efficiency", and "stage"* belong to the semantic context of embodied energy, emphasizing the scientific community's interest in analyzing energy use in the various phases of the construction process [74,75].

### *3.3 Insights into the geographical attention to LCA and Building Sustainability*

As a last step, we gathered information on the geographic orientation of LCA and building sustainability research to see if case studies were focused on specific regions of the world. To achieve this, we created groups of words for each country worldwide – including the corresponding qualifying adjective related to its name. Thus, the "Italy" cluster also includes "Italian" and "Italians". Figure 6 shows the number of papers with case studies and applications related to the different countries, comparing the abstracts resulting from our two search queries.

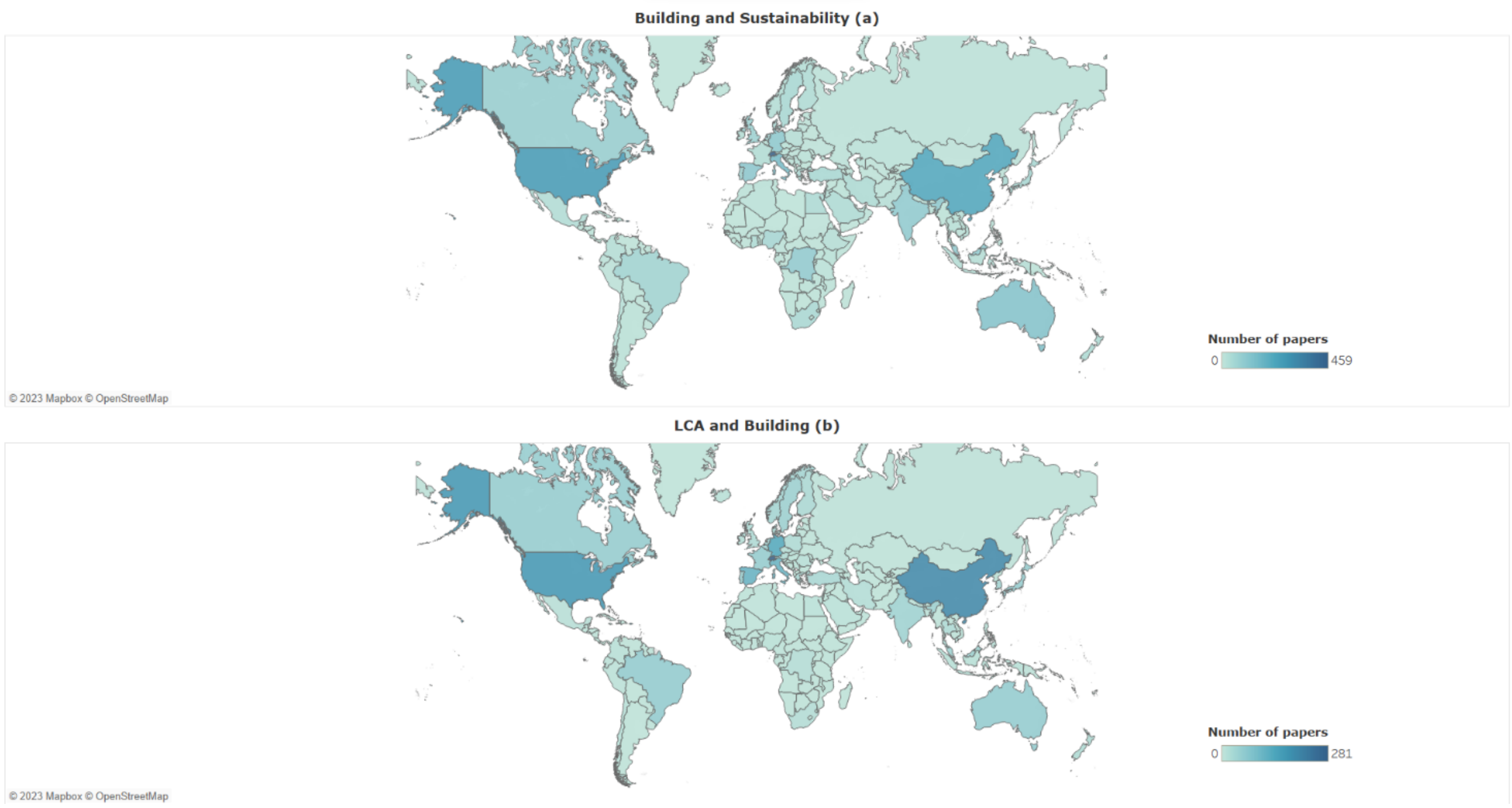

**Figure 6.** Geographic Focus of LCA and Building Sustainability Research

Figure 6 shows a focus on the American continent, followed by Asia, with notable references to China and the United States. In addition, Europe also seems to be mentioned a lot in the abstracts.

The focus on sustainable building aligns with the conclusion drawn in the Global Construction 2025 report [86]. It presents an in-depth analysis of the construction market, projecting a remarkable global growth rate of more than 70% in construction output by 2025. The report indicates that approximately two-thirds of new construction activities will be concentrated in China, India, and the United States. These are the same countries that, in this research, exhibit the most significant interest in the scientific aspects of sustainable building practices. This observation paves the way for future studies exploring the relationship between scientific publications and investment decisions in a country. Such investigations can shed light on how these factors influence each other.

## 4. Conclusions

This review of the literature unveils the evolution of building and sustainability research in four distinct phases. The SBS analysis in Figure 3 illustrates the dynamic changes in the research focus over time. In the early development phase, the surge of interest in *"efficiency"* and *"saving*s" signals a burgeoning emphasis on energy efficiency and resource conservation within building practices. Subsequent phases demonstrate a noticeable increase in the semantic importance of *"materials"*, indicative of a larger concern for sustainable materials and environmental impact assessments. In particular, global initiatives such as the Paris Agreement and the 2030 Agenda for Sustainable Development have the potential to shape recent publication trends in this field. Furthermore, the unexpected emergence of the Covid-19 pandemic appears to have intensified the attention to sustainability and green building practices, as evidenced by the peak in publications during the two years from 2020 to 2022.

Concerning the focus on LCA, the analysis has identified six primary discourse topics in the context of building construction. These topics include "Energy Efficiency", "Materials and Construction Techniques", "Environmental, Social and Economic Impact", "Life Cycle Assessment", "Urban Building Stock", and "Heating Systems". Each topic embodies different research domains within the LCA field, underscoring its multifaceted nature. In these domains, there appears to be a lack of publications that evaluate the social impact that an LCA approach can generate. In addition, aspects of public governance and institutions seem to be missing, resulting in limited interest in these areas. These findings suggest that a more comprehensive and holistic approach to LCA in the building sectors should be adopted, incorporating social and governmental dimensions alongside existing environmental and economic considerations. Such an approach would provide a complete understanding of LCA's impacts and potential improvements, ultimately contributing to more informed and effective sustainability strategies. Simultaneously, the focus on energy savings and efficient resource management is expected to expand further. Considering the growing role of artificial intelligence in the domestic sphere, these advanced energy technologies designed to provide economic services and energy savings are projected to enhance the quality of the life cycle by describing the directions for future research in this field.

The semi-supervised methodology adopted in this work allowed the analysis of a large number of abstracts, covering a wide time span in the literature on sustainability and construction, from 1996 to 2022. This sets this study apart from other reviews focused on narrower time frames

The innovative approach of this methodology also lies in its unique structuring of the research query. By utilizing a concise set of terms, a vast number of abstracts were thoroughly analyzed, resulting in one of the most comprehensive studies of abstracts within the field. This method allowed for a broad and global analysis of abstracts in the construction sector related to Life Cycle Assessment (LCA).

Unlike approaches that rely on predetermined variables chosen by the authors, this research method aimed to include a diverse range of topics within the abstracts, ensuring a more complete sample. Each topic was then carefully evaluated to determine their semantic importance.

Researchers and practitioners in the field seeking to analyze LCA within the construction sector can benefit greatly from this approach. It provides a comprehensive and unbiased view of the topic, free from the influence of discriminatory variables chosen a priori by authors.

In conclusion, this literature review presents researchers with a comprehensive overview of the prominent research trends in LCA and building. This perspective can serve as a valuable tool for envisioning the future and contributing significantly towards achieving the goals outlined in the Paris Agreement and in the UN Agenda 2030.